\documentclass[11pt]{article}
\usepackage{hyperref}
\pdfoutput=1
\begin{document}
\title{Complex dynamics of evaporation-driven convection in liquid layers}
\author{F. Chauvet, S. Dehaeck, P. Colinet \\
\\\vspace{6pt} TIPs (Transfers, Interfaces and Processes), \\Universit\'e Libre de Bruxelles, Belgium}
\maketitle
\begin{abstract}
The spontaneous convective patterns induced by evaporation of a pure liquid layer are studied experimentally. A volatile liquid layer placed in a cylindrical container is left free to evaporate into air at rest under ambient conditions. The liquid/gas interface of the evaporating liquid layer is visualized using an infrared (IR) camera. The phenomenology of the observed convective patterns is qualitatively analysed, 
showing in particular that the latter can be quite complex especially at moderate liquid thicknesses. Attention is also paid to the 
influence of the container diameter on the observed patterns sequence.
\end{abstract}
\section*{}
During the evaporation of a pure liquid layer into dry air the liquid/gas interface is cooled because of the energy consumption for the phase change. This temperature difference across the liquid layer can generate surface-tension-driven convection and/or buoyancy-driven convection in the liquid, depending on the layer thickness. Stable and homogeneous convective patterns have been observed in evaporating liquid layers without heating in [1], for not too high layer thicknesses. Here, the focus is on the complex convective motions which appear when the layer thickness is higher than typically 0.5 mm in our experimental set-up.

In practice, a small amount of volatile liquid (HFE-7100 from 3M) is initially poured in a cylindrical container to form the liquid layer. The container is made of a plexiglas cylinder glued by silicone to an aluminium block. The height of the cylinder is 2 cm and two inner diameters have been tested, namely 59 mm and 34 mm. The liquid volume injected initially is adjusted such as to start with a 3 mm liquid layer thickness. Then the liquid is just left free to evaporate into air at rest under ambient conditions. 
In these conditions, the evaporation process is limited by diffusion of vapor into air and the evaporation rate stays almost constant until the layer is too thin and begins to dewets the aluminium. This has been verified by liquid weight measurements as a function of time, using a precision balance. This can be simply explained by the fact that the diffusive resistance for vapor mass transfer ($\propto$ container height) does not significantly depend on the variation of the small layer thickness. As a consequence, the layer thickness decreases almost linearly with time.

As convection in the liquid is necessarily associated with temperature variations within the liquid (and in particular at the interface), it is chosen here to use an IR camera to follow the time evolution of the temperature distribution, representative of the pattern organization. The IR camera used is a 16-bit Focal Plane Array camera-type (Thermosensorik, InSb 640 SM) with a spectral band sensitivity ranging from 1.0 to 4.8 $\mu$m and a 640x512 pixels$^2$ sensor cooled at 77 K by a Stirling engine.
As the liquid layer is semi-transparent, in the spectral band sensitivity of the IR camera specific work has to be done to convert 
the recorded IR signal in temperature. Here, the IR camera is just used as a visualization tool to analyze the spatio-temporal dynamics of convective motions in the liquid layer.

Evaporation experiments have been done for the two container inner diameters, 59 mm and 34 mm. In both cases convection appears immediately after the injection of the liquid. In the 59 mm diameter container, fluid motion is complex but we can identify some individual large-scale structures which are quite stable. As the layer thickness decreases because of the evaporation, these large structures are more and more organized, their number increases and their size decreases. Progressively, the pattern consists of stable convective cells which split
into smaller ones. During this period, the contrast between cells decreases on the recorded IR images until there is no visible convection. Finally, the layer thickness continues to decrease without convection motion until the dewetting sequence starts.

In the 34 mm diameter container, the phenomenology is different. No individual large-scale structures are observed as it is the case in the 59 mm diameter container. The pattern is globally axisymetric with a cold point at the center of the container. After a while, convective cells are suddenly created from the container center and ejected towards the container wall. Contrary to all other patterns transitions seen here, this transition can be particularly well identified. After this violent event, the pattern looks like to the pattern observed in the 59 mm diameter container with more and more organized structures, displaying similar patterns sequences.

It can be concluded that, for layer thicknesses investigated here, chaotic convective patterns depend on the container diameter when the convective structures are large compared to the container diameter (strong spatial confinement). At smaller depths and hence smaller pattern wavelengths, the dynamics is much less affected by the container walls (weak spatial confinement). In the former case, a particular symmetry-breaking transition has been identified, which will be the subject of further investigations.
\section*{Acknowledgments}
Supported by the Marie Curie MULTIFLOW Network,  by ESA \& BELSPO PRODEX projects, and by FRS - FNRS

\section*{References}
[1] H. Mancini, D. Maza, {Pattern formation without heating in an evaporative convection experiment, Europhys.       Lett., 66 (6), 812–-818, 2004}
\end{document}